\documentstyle[preprint,prb,aps]{revtex}
\begin{document}
\draft
\tighten
\title
{Flux cutting in YBa$_{2}$Cu$_{3}$O$_{7-\delta}$ single crystals:
experiment and phenomenological model.}

\author
{D. L\'opez, G. Nieva\cite{conicet}, and F. de la Cruz}
\address{Comisi\'on Nacional de Energ\'{\i}a At\'omica,
Centro At\'omico Bariloche, 8400 S. C. de Bariloche,
Argentina}
\author{Henrik Jeldtoft Jensen and Dominic O'Kane}
\address{Deparment of Mathematics, Imperial College, London SW7 2BZ,
United Kingdom}

\maketitle

\begin{abstract}
We measured the current induced loss of vortex correlation in YBCO crystals
using the pseudo DC flux transformer geometry.
We find that the current $I_{cut}$
at which the top and bottom voltage drops differ
is a linear function of temperature independent of pinning.
Although the vortex correlation length at $I_{cut}$ coincides with the sample
thickness, the experimental data show that $I_{cut}$
is sample thickness independent.
The observed behavior with temperature, magnetic field and thickness is
described
by a phenomenological model.
\end{abstract}

The laminar structure of the high temperature superconductors has
important influence on the vortex behavior in the mixed state. This laminar
structure and the presence of thermal fluctuations yields an extremely rich
$H-T$ phase diagram.\cite{bishop} Recent experiments\cite{nolocal},
using a DC flux transformer contact configuration, have shown that the
correlated vortex motion can be destroyed by either thermal fluctuations or
the application of large driving forces.
The non-homogeneous current distribution induced in electrical transport
measurements in YBCO(123) single crystals, in the
DC-transformer\cite{giaever} configuration,
has been used\cite{nolocal,philos,out} to determine the correlation
length of the vortices in the direction of the applied field
(c crystallographic direction). It was shown\cite{nolocal,philos}
that there is a well defined temperature, $T_{th}(H)$, above
the irreversibility line $T_{i}(H)$, where the correlation
of the vortex velocity across the sample is lost. This has been
interpreted \cite{nolocal,philos} as a decoupling of the vortices
nucleated in the Cu-O planes. Since this decoupling takes place in
the linear current-voltage response regime it is
concluded \cite{nolocal,philos} that at that temperature the vortex
correlation length in the direction of the applied field coincides
with the thickness, $d$, of the sample. As a consequence the transport
measurements in samples of different thicknesses have been used \cite{out} to
determine the temperature dependence of the correlation length in the
c direction at different fields. On the other hand transport measurements
in the linear regime using a Montgomery type analysis have led to the
conclusion\cite{nolocal} that the transport properties of the vortex
liquid phase has a non-local character.\cite{huse} The non-local character
of the electrodynamics induces an enhancement of the apparent anisotropy
of the resistance of the vortex liquid in the DC transformer
configuration.\cite{nolocal}

In this paper we concentrate our interest in the study of the
I-V characteristics for $T < T_{th}$ and temperatures above and below
the reversibility line, defined as the temperature $T_{i}(H)$ where
the resistance of the $ab$ planes tends to zero. For $T > T_{i}(H)$ the
I-V characteristics show\cite{ffh} a linear regime at low currents
and
become non linear at higher currents. For $T <T_{i}(H)$ the response is
non linear\cite{ffh} for all currents. It is shown that for
$T < T_{th}$ flux
cutting is induced by a current $J_{cut}(H,T,d)$, in the non-linear
response regime.

The experimental results are consistent with a current distribution flowing
near the sample surface.
This allowed us to extend the model developed by
J. Clem {\em et al.}\cite{clem} to explain our results obtained using a
non-homogeneous current distribution.
The results of the  model provide a qualitative
comprehension of the field, temperature and sample
thickness dependence of $J_{cut}(H,T,d)$. It also shows that the
understanding of the thickness dependence\cite{nolocal,out}
of $T_{th}$ goes beyond
the assumptions made by the model and requires the analysis of
the contribution of thermal fluctuations in the vortex transport
properties.

The YBCO single crystal was prepared as indicated in Ref. 5. The results
presented in this paper were obtained using the same samples and
contact configuration described in Refs. 5 and 6.

In Fig. 1 we plot typical I-V characteristics for temperatures
above (1a) and below (1b) $T_{i}=89.7K$ at a field of $10kOe$.
In this case the sample thickness is $d=20 \mu m$. The experiment was
made injecting current through contacts 1 and 4, $I_{14}$, and measuring the
voltages $V_{23}$ and $V_{67}$, see insert in Fig. 1. At low currents
$V_{23} = V_{67}$ indicating that the vortices move correlated
across the sample. Increasing the driving current, $V_{23}$ becomes different
from $V_{67}$ at a well defined current $I_{cut}$ (arrows in Fig. 1) where
flux cutting is induced.

Figure 2a shows the cutting current density defined as
$J_{cut}=I_{cut}/(d \times w)$ (where $w$ is the  width
of the sample) as a function of temperature for
samples of different thickness. The data show that $J_{cut}$ decrease
linearly with temperature vanishing at $T_{th}$.\cite{nolocal,rodri}
For a given temperature, $J_{cut}$ decreases when increasing the thickness.
In Fig. 2b we show the
same data but scaled by the sample thickness. It is clearly seen that
$J_{cut} \times d$
becomes thickness independent. This result indicates that flux cutting is
induced by a force that can be associated with a surface current density
given by $I_{cut}/w$.
That is, the transport current should be considered as flowing in a
region close to the surface in a thickness much smaller than d, in
agreement with the suggestion made in ref. 7 and with the
experimental results obtained in the linear response regime.\cite{nolocal}

It is interesting to point out that the data above and below $T_{i}(H)$
fall on the same straight line. This is demonstrated in Fig. 2a where
$T_{i}$ ($T_{i}/T_{th} = 0.985$) is indicated by the arrow for the
sample with $d=48 \mu m$.
Thus, flux cutting should be related to the properties
of the flux structure, independently of the presence of effective
pinning centers. That is, the dynamic vortex correlation length in the
field direction is not modified by the introduction of pinning.

With the results and considerations made before we present
now a simple model which enables to rationalize the
observed
field and temperature dependence of I$_{cut}$. The model also explains
why I$%
_{cut}$ is independent of pinning whereas the voltage V$_{cut}$ at the
cutting does depend on pinning.

Consider a stack of N superconducting layers with vortex pancakes
moving in these layers. The equation of motion for vortex number $i$ in
layer number $\nu $ (see Fig. 3a) is given by:
\begin{equation}
\eta v_{i,\nu} = \delta _{1,\nu} f - f_{i,\nu}^p+ f_{i,\nu}^\nu+ %
\left( 1-\delta _{1,\nu}\right) f_{i,\nu}^{\nu - 1}-\left( 1-\delta _{N,\nu}
\right) f_{i,\nu}^{\nu + 1}
\label{eq1}
\end{equation}
where $\eta $ is a friction coefficient which for convenience we put equal to
$1$.
The terms $f_{i,\nu}^p$ denote the total pinning force on vortex $(i,\nu)$.
The term $f_{i,\nu}^v$ is the force from all the vortices in layer $\nu$ on
vortex $(i,\nu)$ and $f_{i,\nu}^{\nu \pm 1}$ denotes the interlayer
forces between the vortices.
We assume that the effect of the external current is to give rise to a
driving force $f$, which is confined to the top layer. This assumption is
consistent with the experimental observation as discussed above. It is also
in accordance with the phenomenological viscous hydrodynamics of Huse and
Majumdar.\cite{huse}
However, it should be mentioned that Huse and Majumdar work in the limit of
linear resistance, whereas the cutting by current always occur in the
non-linear region.

We want to determine the average velocity  in layer $\nu$:
$v _{\nu} =\left\langle
v_{i,\nu}\right\rangle $, since the voltage drop over that layer is
proportional to $v_{\nu}$. We consider the situation where the average velocity
is the same in each layer, i.e., $v = v_{1}, v_{2},......v_{N}$,
thus $V_{top}=V_{bottom} (V_{23}=V_{67})$. Assuming that the average
of the pinning force over
vortices and time is independent of the layer number, we obtain the
following equation:

\begin{equation}
v=\delta _{1,\nu}f-f^p+\left( 1-\delta _{1,\nu}\right) f_{\nu}^{\nu - 1}-
\left(1-\delta _{N,\nu}\right) f_{\nu}^{\nu + 1}\\
\label{eq2}
\end{equation}

\noindent
This set of equations is readily solved to give:

\begin{equation}
f_{\nu}^{\nu+1}=\frac{N-\nu}N f
\label{eq3}
\end{equation}

The largest stress is between the two top layers. The driving force at which
the velocities in layer 1 and 2 starts to differ is given by:

\begin{equation}
f_{cut}=\frac N{N-1}f_{1}^{2, max }\simeq f^{max }
\label{eq4}
\end{equation}
where $f^{max }$ is the maximum force which the bonds between pancakes in
adjacent layers can sustain. Notice that $f_{cut}$ is independent of the
pinning forces. In contrast, the velocity (or voltage) at cutting
does depend on the pinning. By substituting Eq. 4 into the equation for
the average velocity one obtain $v_{cut}= f^{max}/(N-1) - f^p$.
Thus if one knows the friction coefficient $\eta$ one can determine
the intrinsic property $f^{max}$ together with the extrinsic
pinning force $f^p$ by a combined measurement of the voltage and
the current at cutting.

To estimate $f^{max }$ is like to estimate the yield strength of a
material. It is not easy. Let $\Delta x_{max }$ denote the displacement of
vortices in layer $2$ at which plastic slip sets in (see Fig. 3b and 3c).
Let $\Delta E$ denote
the associated increase in the energy due to the deformation. We estimate
\begin{equation}
f^{max } = \frac{\Delta E}{\Delta x_{max }}
\label{eq5}
\end{equation}

Slip will occur when $\Delta x_{max }$ reaches some fraction of $a_o$, the
distance between the vortices. For small $a_o$ (Fig. 3b), i. e.
large magnetic field, we
expect $\Delta E=\kappa _1\Delta x^2$ where $\kappa _1$ will be a short wave
length tilt modulus. For lower fields (Fig. 3c), $a_o$ increases and so do
$\Delta
x_{max }$. In this case  we should rather connect $\Delta E$ with the energy
of the Josephson
vortex running in between the planes connecting  the  vortices  in
the two layers.
This suggest that $\Delta E$ becomes linear  in  $\Delta  x$
for low  fields:  $\Delta  E  =  \kappa_2  \Delta  x$.  Both
constants $\kappa_1$ and $\kappa_2$ will scale with the superfluid
density $|\psi |^2$. The maximum force will behave like

\begin{equation}
f_{max} \simeq |\psi |^2 G(B),
\label{eq6}
\end{equation}

\noindent
where  $G(B)   =   1$   for   low   fields   and   $G(B)   \simeq
B^{-1/2}$ for high fields.

We    have    $|\psi    |^2    \simeq    (1-\frac{T}{T_c})(1     -
\frac{B}{B_{c2}(T)})$.\cite{brandt} Furthermore,
$f_{max} = \phi_o  J_{cut}  \frac{l}{c}$,
where $l$ denotes the thickness  of  the  layer ($l \ll d$)   that  carries
the
transport current which gives rise to  the  Lorentz  force.  In the model
$J_{cut} = I_{cut}/(l \times w)$, therefore

\begin{equation}
I_{cut} \simeq (1 - \frac{T}{T_c}) (1 - \frac{B}{B_{c2}(T)}) G(B),
\label{eq7}
\end{equation}

\noindent
independent of $l$.

Using $B_{c2}(T) = (1 - \frac{T}{T_c})B_{c2}(0)$ we finally obtain

\begin{equation}
I_{cut} \simeq (1 - \frac{T}{T_c} - \frac{B}{B_{c2}(0)}) G(B).
\label{eq8}
\end{equation}

\noindent
According to this expression, $I_{cut}$ depends  linearly  on  the
temperature. As B is increased the  intersection  of  $I_{cut}(T)$
with the {\em x}-axis decreases linearly  in  $B$.  The  slope  of
$I_{cut}(T)$ is independent of $B$ for small $B$.  For  large  $B$
the slope decreases as $B^{-1/2}$.
This is in qualitative agreement  with  the  experimental  results  shown  in
the inset of Fig. 2b. For a quantitative analysis of the results, more data
at higher fields are necessary.

As was discussed above, the experimental results shown in Fig. 2a prove
that $I_{cut}$ is independent of pinning. This agrees with our model, as
shown by expression (\ref {eq4}), where $f_{cut}$ is found to be pinning
independent despite pinning is included in the equations of motion
(see equation
\ref{eq1}). Thus, expression (\ref{eq8}) describes the experimental
temperature and field dependence
of $I_{cut}$ and shows that $I_{cut}$ is an intrinsic
property of the flux structure, in agreement with the data of Fig. 2a.

We have restricted
our  discussion  to  purely  mean  field
considerations neglecting thermally induced vortex fluctuations.
We now want to turn our  attention  to  the  field  dependence  of
$T_{th}$. As shown in Fig. 4 one finds experimentally that $T_{th}$
depends linearly on the magnetic field.
In accordance with the definition of $T_{th}$, it appears natural
to associate  this  temperature  with
the temperature where vortex-antivortex pairs thermally  activated
in the superconducting planes unbind. The unbinding in zero  field
happens when the ratio
$T/|\psi  |^2$ achieves  a  certain  invariant  value\cite{minnhagen}:

\begin{equation}
T_{ub}/ |\psi(T_{ub})|^2 = const.
\label{eq11}
\end{equation}

\noindent
It has previously been suggested
\cite{Jensen} that the effect of the external  magnetic  field  on
the unbinding transition in a first approximation can  be  obtained
simply by including the mean field dependence on $B$ in $|\psi|^2$
in Eq. (\ref{eq11}).

This leads to

\begin{equation}
\frac{T_{ub}(B)}{T_{ub}(0)} = 1-\frac{B}{B_{c2}(T=0)}.
\label{eq12}
\end{equation}

\noindent

The linear field dependence of $T_{ub}$ and the decoupling of  the
layers\cite{Weber} at $T_{ub}$, strongly suggest that $T_{ub}$ is identical to
$T_{th}$. The dependence of $T_{th}$ on thickness arises  then  as
an effect  of  the  correlation  length  (in  the  field  direction)
connected with the unbinding  transition  becoming  equal  to  the
sample thickness.\cite{nolocal}

We have been able to show that flux cutting in YBCO(123),
using the DC transformer configuration, is induced by currents
flowing at the sample surface, in strong agreement with the
suggestion made in ref. 7 and the previous results of ref. 2,
obtained in the linear response regime. This has allowed us to
formulate a simple model which disregarding the effect of thermal
fluctuations, describes qualitatively the results in the non-linear
response regime.
The temperature dependence is correctly described
by the model, assuming that the coupling between planes has its
origin in a Josephson energy.

We thank E. Jagla, C. Balseiro and E. Osquiguil for discussions
in the course of preparing this manuscript.
This work was supported by The British Council - Fundaci\'on Antorchas
and by the British EPSRC.

\begin{figure}
\caption{$V_{23}$ and $V_{67}$ as a function of driving current $I_{14}$ for
two different temperatures: a) 89.9K and b)89.4K. Also shown is $\Delta V=
V_{23}-V_{67}$. The sample geometry is shown in the inset. The arrows show
the current, $I_{cut}$, where $V_{23}$ and $V_{67}$ start to differs.}
\label{f1}
\end{figure}

\begin{figure}
\caption{ a) Temperature dependence of $J_{cut} = I_{cut}/(d \times w)$ for
different sample thickness, for an applied field of $10 kOe$. The arrow shows
the corresponding $T_{i}$ for the sample with $d=48 \mu m$. The solid
lines are linear fit to the data.
b) Same data of Fig. 2a scaled by the sample thickness $d$:
($J_{cut} \times d$)
as a function of temperature. In the inset is shown the temperature and
field dependence of $I_{cut}$ for the sample with $d=32 \mu m$. The data
were fitted using expression (\protect\ref{eq8})(solid lines).}
\label{f2}
\end{figure}

\begin{figure}
\caption{Schematic diagram showing the velocity of the
pancakes vortices in different superconducting layers (see text).
Sketch of the vortex configuration in adjacent layers for large (b)
and small (c) fields. $\Delta x$ is the relative displacement between
vortices in adjacent layers due to the external current.}
\label{f3}
\end{figure}

\begin{figure}
\caption{Phase diagram showing the temperature $T_{i}$ and $T_{th}$ for
a sample with $d=20 \mu m$. The solid line is a linear fit and
the dotted one is a guide to the eye.}
\label{f4}
\end{figure}
\end{document}